\documentclass[preprint,eqsecnum,aps,showpacs]{revtex4}
\usepackage{dcolumn}
\usepackage{graphicx}
\usepackage{amsmath}
\usepackage{amssymb}
\usepackage{epsfig}
\usepackage{float}
\usepackage{latexsym}
\usepackage{rotating}
\begin{document} 
\title{Kinematics of flows on curved, deformable media}

\author{
Anirvan Dasgupta\footnote{Electronic address: {\em anir@mech.iitkgp.ernet.in}}${}^{}$}
\affiliation{Department of Mechanical Engineering and Centre for Theoretical Studies  \\
Indian Institute of Technology, Kharagpur 721 302, India}
\author{
Hemwati Nandan\footnote{Electronic address: {\em hnandan@cts.iitkgp.ernet.in}}${}^{}$}
\affiliation{Centre for Theoretical Studies  \\
Indian Institute of Technology, Kharagpur 721 302, India}
\author{
Sayan Kar\footnote{Electronic address: {\em sayan@cts.iitkgp.ernet.in}}${}^{}$}
\affiliation{Department of Physics and Centre for Theoretical Studies  \\
Indian Institute of Technology, Kharagpur 721 302, India}
\begin{abstract}
Kinematics of geodesic flows on specific, two dimensional, curved surfaces
(the sphere, hyperbolic space and the torus) are
investigated by explicitly solving the evolution 
(Raychaudhuri) equations for the expansion, shear and rotation, for a variety
of initial conditions. For flows on the sphere and on hyperbolic space, 
we show the existence of singular (within a finite
value of the time parameter) as well as non-singular solutions. We illustrate
our results through a {\em phase diagram} which demonstrates 
under which initial conditions (or combinations thereof) we end up
with a singularity in the congruence and when, if at all,
we can obtain non--singular solutions for the kinematic variables. 
Our analysis portrays the differences which arise
due to positive or negative curvature and also explores the
role of rotation in controlling singular behaviour. 
Subsequently, we move on to geodesic flows on two dimensional spaces with 
varying curvature. As an example, we discuss flows on a torus. 
Characteristic oscillatory features, dependent on the ratio of the two
radii of the torus, emerge in the solutions for the expansion, shear and rotation.  Singular (within a finite time) and non--singular behaviour of the 
solutions are also discussed.
Finally, we conclude with a generalisation to three dimensional spaces of
constant curvature,
a summary of some of the generic features obtained 
and a comparison of our results with those for flows
in flat space.\\

\end{abstract}

\maketitle

\section{Introduction}

Geodesic flows in curved spaces have been a topic of interest among
mathematicians for many years. However, as far as we are aware,
 a detailed study of
the kinematics of such flows, does not exist in the literature.
The kinematics of flows can either be linked with the evolution of particle distribution on a manifold, or with the evolution of deformations 
of a medium itself. This aspect may be understood as follows. 
\begin{figure}[H]
\begin{center}
\includegraphics[scale=0.7]{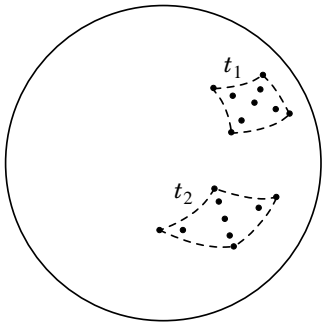}
\end{center}
\caption{Particle distributions at two time instants $t_1$ and $t_2$ evolving under (geodesic) motion on a spherical surface.}
\label{fig1a}
\end{figure}
\noindent Imagine a spherical surface on which we have particles sprinkled randomly as shown in Fig.~\ref{fig1a}. 
Given some initial motion, these particles will move along trajectories 
governed by the equations of motion (geodesics, say, as a special case). 
In an active viewpoint, we might look at a region on the sphere containing a 
certain number of particles at an initial time and ask-- do the particles 
accrete, spread out or redistribute themselves in an area-preserving way,
as time evolves? The same question may also be posed for understanding
deformations of a medium as well, where the particles now imply the points of
the medium. Such investigations in a real four dimensional scenario, are of fundamental importance in studying
accretion near black hole horizons, structure formation and evolution of the
universe at different scales, and motion of fluid flows to name a few. \\

In \cite{adg}, we restricted ourselves to flows in flat space
where the evolution of each point on the medium was along
a curve which solved the Newton's second law with damping and
elastic forces added. What happens if we have
flows on surfaces of nontrivial curvature? To this end, here, we
look at geodesic flows on a sphere, a saddle surface as well as a surface of 
varying curvature-- the torus, and figure out
the kinematics of flows on them subject to
specific initial conditions.
Curvature creates non--trivial effects on the flows, which need to be
quantified. We are fortunate that in some cases, the set of equations 
governing the behaviour of flows can
indeed be solved analytically. Therefore, we are able to see explicitly
how the kinematical quantities evolve. Where analytical solutions are
impossible, we resort to numerical methods in order to arrive at
our conclusions. 

The kinematics of flows on deformable media is 
characterised by the kinematical quantities-- 
the isotropic expansion, shear and rotation or vorticity \cite{ll}
(henceforth referred as ESR). 
The deformation of the medium (or a local region in the medium) 
is related to the behaviour, with time, of the congruence of geodesics 
and can be characterised in terms of the time evolution of some 
deformation vector \cite{toolkit,adg}. The analysis of this evolution 
ultimately leads to a local, first order, nonlinear and coupled
set of equations for the ESR variables, known in the literature, as the 
Raychaudhuri equations \cite{review}.

The Raychaudhuri equations are well-known 
in the study of spacetime singularities in gravitation
and cosmology \cite{hawk}-\cite{ciufolani}. An important 
consequence which emerges from the analysis of the equation
for the expansion 
is the focusing of  geodesics \cite{Kar}. 
Moreover, the use of Raychaudhuri equations in the proofs of singularity 
theorems \cite{Penro,Haw} (through the notion of focusing of geodesics)
is, today, a well-established fact \cite{wald,joshi}. 

An important assumption about our forthcoming analysis may be noted here.
In \cite{adg}, we assumed that the medium 
(more precisely, the particles that make the medium) does not develop
curvature due to the deformation (i.e the flat space remains flat).
Here too,  we assume that the deformations do not change the curvature of 
the medium.

Our plan in this paper is as follows.
We first derive the Raychaudhuri equations in two dimensional 
curved spaces of constant curvature. We then
recall the exact solutions of the geodesic equations in the spherical and 
hyperbolic geometry, which are required as inputs while solving the
Raychaudhuri equations. The appearance of a
finite time singularity depending on the initial conditions on the expansion, 
shear and rotation variables is  investigated analytically 
by including the effects of stiffness and viscous damping. 
The generic features of the solutions for 
different values of the parameters appearing in the Raychaudhuri equations 
and the initial conditions are clearly brought out and summarized. 
Some specific cases are also solved numerically and discussed in detail.
The equations presented in this two dimensional case can easily be
extended to
curved higher dimensional spacetimes with space-like sections of constant 
curvature
(maximally symmetric spacetimes). This aspect of our work here is one of our
motivating factors because, in future, we intend to carry out
similar analysis in spacetimes which represent specific
gravitational fields and arise as solutions in General Relativity.  

Following the analysis of flows on constant curvature spaces, we move
on to spaces of varying curvature. We discuss the specific example of
flows on a torus for various classes of geodesics on it and comment
on particular features in the solutions for the expansion, shear
and rotation. 

Finally, in the last section, we offer our comments and conclusions
as well as our views on possible future work. 

\section{Geodesic flows on two dimensional spaces with constant curvature}

\subsection{Expansion, rotation and shear}

The deformations in a two dimensional deformable medium can be characterised 
in terms of the time evolution of the  deformation vector $\xi^{i}$ (where $i=1,2$) which connects two infinitesimally separated points of the medium at any time instant \cite{toolkit,adg}. The points at a later time instant move an infinitesimal distance along the respective instantaneous velocity directions \cite{adg}. For the purpose, the time evolution of the deformation vector for short time intervals  can be given as,
\begin{equation}
\frac{d\xi^{i}}{dt} = B^i_{\,\,j} (t)\, \xi^{j}+{\cal O}(\Delta t^2),
\label{devveq}
\end{equation}
where the time dependent second rank tensor $B^{i}_{\,\,j}(t)$ governs the dynamics of the deformable medium. Further, using the equation (\ref{devveq}), the second order derivative of  the deformation vector ${\xi}^i$ with respect to time can be calculated as follows,
\begin{equation}
\left (\frac{dB^i_{\,\,j}}{dt} + B^i_{\,\,k}B^k_{\,\,j} 
\right ) \xi^j={\ddot\xi}^i.
\label{ddotxi}
\end{equation}
In order to describe the congruence's behaviour,  the evolution tensor $B^i_{\,\,j}$ in equations (\ref{devveq}) and (\ref{ddotxi}) can be decomposed as the linear  combination of expansion, shear and 
rotation in the following form,
\begin{equation}
B^i_{\,\,j} = \frac{1}{2}{\theta} \,\delta^i_{\,\,j}  +\sigma^i_{\,\,j} + \omega^i_{\,\,j}, \label{bij} 
\end{equation}
where $\theta$, $\sigma^i_{\,\,j} $ and $\omega^i_{\,\,j}$ represent, respectively, the 
expansion, shear and rotation. The kinematics of deformations can now  quantified in terms of expansion, shear and rotation (ESR) of the medium. The equation (\ref{bij}) can be written in the following $2\times 2$ matrix form,
\begin{equation}
B^i_{j} =
\left(
\begin{array}{cc}
\frac{1}{2}\theta & 0\\
0&  \frac{1}{2}\theta\\
\end{array} \right)
 + \left(
\begin{array}{cc}
\sigma_{+} & \sigma_{\times} \\
\sigma_{\times} & -\sigma_{+} \\
\end{array} \right)
 +
\left(
\begin{array}{cc}
0 & -\omega \\
\omega& 0 \\
\end{array} \right),
 \label{ctrans}
\end{equation}
where $\sigma_+$ and $\sigma_\times$ represent the shear parameters while $\omega$ is the rotation parameter which along with the expansion scalar $\theta$, characterise the deformations of the medium \cite{toolkit,adg}.
\subsection{The evolution equations} The evolution equations for the 
congruence of geodesics consist of the Raychaudhuri equations for the ESR 
variables and the geodesic equations (corresponding to the metric of interest) 
which are presented below. 

\subsubsection{Raychaudhuri equations}

In order to have the evolution (Raychaudhuri) equations for  ESR variable, we need to consider a general expression for $\ddot \xi^i$ is as follows, 
\begin{equation}
\ddot \xi^{i}\, = - R^{i} _{\,\, ljm} u^{l}u^{m}\xi^{j} - K^i_{\,\,j} \, \xi^j -\beta \, \dot{\xi}^i\, , \\
\label{sdform}
\end{equation}
where $K^i_{\,\,j}$ and $\beta$ are the stiffness and viscous damping
parameters in the medium, respectively. It may be noted that the first term on the right hand side of equation (\ref{sdform}) is clearly because of the curvature contribution and is absent while dealing with the case of flat space \cite{adg}. The  $K^i_{\,\,j}$ in equation (\ref{sdform}) has the form given as,
\begin{equation}
K^i_{\,\,j} =
\left(
\begin{array}{cc}
k+k_{+}& k_{\times} \\
k_{\times} & k-k_{+} \\
\end{array} \right).
\end{equation}
Using equations (\ref{ddotxi}) and (\ref{sdform}) lead to,
\begin{equation}
\dot{B}^i_{\,\,j}+B^{i}_{\,\,k}B^k_{\,\,j} + K^i_{\,\,j}+\beta B^i_{\,\,j}= -  R^{i} _{\,\, ljm} u^{l}u^{m}.
\label{dotb}
\end{equation}
In order to derive the evolution equations for ESR variables, let us consider a  two dimensional metric in the following form,
\begin{equation}
ds^2 = \frac{ dr^2}{ 1-\kappa r^2} \,  +  r^2 d \varphi^2 ,
\label{metric}
\end{equation}
where $\kappa = 0, +1, -1$ denotes the curvature corresponding to the flat, spherical and hyperbolic geometries, respectively. Finally, using the decomposition of evolution tensor (\ref{bij}) in the equation (\ref{dotb}), one obtains the
 Raychaudhuri equations for the ESR variables as follows,  
\begin{equation}
\dot \theta + \frac{1}{2}{\theta^2}+\beta \theta  + 2 (\sigma_{+}^{2} + \sigma_{\times}^{2} -\omega^2) + \kappa + 2k =0,
\label{theta}
\end{equation}
\begin{equation}
\dot \sigma_{+}+ (\beta +\theta) \, \sigma_{+} -\frac{\kappa}{2} + k_{+} =0,
\label{sigma+}
\end{equation}
\begin{equation}
\dot \sigma_{\times}+ (\beta +\theta) \, \sigma_{\times}  + k_{\times}=0,
\label{sigmacross0}
\end{equation}
\begin{equation}
\dot \omega+ (\beta + \theta)\, \omega  =0
.\label{omega}
\end{equation}
For $\kappa =0$, these equations exactly reduce to the case of flat space time \cite{adg}. It may be recalled that $R_{ij}=g_{ij} R/2$ in two dimensions. The analytical solutions of the equations (\ref{theta})-(\ref{omega}) can be found by solving the geodesics equations corresponding to the metric (\ref{metric}). 

\subsubsection{Geodesic equations}

The geodesic equations derived from the metric (\ref{metric}) are as follows,
\begin{equation}
\ddot r + \frac{\kappa r}{(1-\kappa r^2)} \dot r^2 - r\, (1-\kappa r^2)\,  \dot \varphi ^2 = 0,
\label{rg}
\end{equation}
\begin{equation}
r \, \ddot \varphi + {2\dot r} \dot \varphi =0,
\label{tg}
\end{equation} 
where the dot denotes the derivative with respect to an affine  parameter $\lambda$ which we shall identify, without any loss of generality, with the
parameter $t$ quoted earlier. The equation (\ref{tg}) leads to $\dot \varphi = c/r^2$ where $c$ 
is an integration constant. These equations can be solved separately for 
 $\kappa = \pm 1$ which are discussed below.  \\
\noindent (i) \, {\bf \underline {For \mbox{\boldmath $\kappa$} =1}} : \\
Let us consider $r= \sin \phi$, which transforms the equations (\ref{rg}) and (\ref{tg}) as follows,
\begin{equation}
\ddot \phi  - \sin\phi \, \cos\phi\, \, \dot \varphi^2 = 0,
\label{rg1}
\end{equation}
\begin{equation}
\ddot \varphi + 2 \cot\phi \, \, \dot \phi \, \dot \varphi  = 0.
\label{tg1}
\end{equation} 
The equations (\ref{rg1}) and (\ref{tg1}) are satisfied with $\phi = \pi/2$ and $\dot \varphi  = d$ (where $d$ is a constant) which  represents an equatorial great circle.\\
\noindent \,  (ii) {\bf \underline {For \mbox{\boldmath $\kappa$} =-1 }} : \\
With  $r= \sinh \phi$, the equations (\ref{rg}) and (\ref{tg}) transforms in the following form,
\begin{equation}
\ddot \phi  - \sinh\phi \, \cosh\phi \, \, \dot \varphi^2 = 0,
\label{rg2}
\end{equation}
\begin{equation}
\ddot \varphi + 2 \coth\phi \, \, \dot \phi \, \dot \varphi  = 0.
\label{tg2}
\end{equation} 
The equations (\ref{rg2}) and (\ref{tg2}) have a solution  with $\phi = \lambda$ and $\varphi = f$, where $f$ is a constant. \\
In order to evaluate the constants involved in the solutions of the geodesic equations,  the normalisation condition $
g_{\alpha\beta} u^{\alpha}u^{\beta} = 1
$  leads to,
\begin{equation}
\dot r^2 + r^2 \, {(1-\kappa r^2) \, }\dot \varphi^2 = (1-\kappa r^2).
\label{nc2}
\end{equation}
In view of the solution for geodesic equations (\ref{rg1}) and (\ref{tg1}) with  (\ref{nc2}), the constant $c =d= \pm 1$ for the case $\kappa=1$. On the other hand $c=0$ for $\kappa=-1$ for both the solutions considered for the geodesic equations (\ref{rg2}) and (\ref{tg2}). 

It is worth noting that for the corresponding case in $2+1$ dimensions 
with $-dt^2$ added in the line element given in  equation (\ref{metric}), 
one will  obtain an additional geodesic equation corresponding to $t$ as 
$\ddot t =0$.  
Thus, $t= a\lambda +b$ and with the choices $a=1$ and $b=0$,  we
get back the above set of 
the geodesic equations. So, the generic features of geodesics,
will remain same as they were in two  dimensions. 
One may also note in the passing that in the case of cylindrical geometry 
with constant radius $\rho$ represented by the metric
$ds^2=dz^2+\rho^2d\vartheta^2$, the geodesic equations are obtained as 
$\ddot{z}=0$ and $\ddot\vartheta=0$. In 
such a case, the Raychaudhuri equations essentially reduce to those obtained 
for the case of flat (Euclidean) space (see \cite{adg}).

\subsection{Analytical solutions of Raychaudhuri equations}
For the above mentioned choice of solutions of geodesic equations,  
the Raychaudhuri equations (\ref{theta} - \ref{omega}) for ESR variables reduce to the following simplified form,
\begin{equation}
\dot \theta + \frac{1}{2}{\theta^2} +  \beta \theta + 2 (\sigma_{+}^{2} + 
\sigma_{\times}^{2} -\omega^2) + \kappa  + 2 k = 0,
\label{thetak}
\end{equation}
\begin{equation}
\dot \sigma_{+}+ (\beta + \theta) \, \sigma_{+} -p=0,
\label{sigma+k}
\end{equation}
\begin{equation}
\dot \sigma_{\times}+  (\beta + \theta) \, \sigma_{\times}-q = 0,
\label{sigmacross}
\end{equation}
\begin{equation}
\dot \omega + (\beta + \theta) \, \omega =0,
\label{omegak}
\end{equation}
where $p=\kappa/2-k_+$ and $q=-k_\times$. 
The equations (\ref{thetak})-(\ref{omegak}) represent the dynamics of the ESR 
variables.
It may be mentioned here that for flat (Euclidean) geometry,
the evolution equations for the ESR variables (see \cite{adg}) are recovered by putting $\kappa=0$ and
redefining $p=-k_+$.\\
Using the substitution $\theta = \ddot{Z}/\dot{Z}-\beta$ in (\ref{thetak})-(\ref{omegak}) leads to,
\begin{eqnarray}
&&\frac{1}{2}\dddot{Z}\dot{Z}-\frac{1}{4}\ddot{Z}^2+\dot{Z}^2(\sigma_+^2+\sigma_\times^2-\omega^2)+
\left(\kappa+2k-\frac{\beta^2}{2}\right)\frac{\dot{Z}^2}{2}=0,
\label{modeq1} \\
&&\dot{Z}\dot{\sigma}_+ +\ddot{Z}\sigma_+-p\dot{Z}=0, \label{modeq2}\\
&&\dot{Z}\dot{\sigma}_\times+\ddot{Z}\sigma_\times-q\dot{Z}=0, \label{modeq3}\\
&&\dot{Z}\dot{\omega}+\ddot{Z}\omega=0. \label{modeq4}
\end{eqnarray}
Solving for $\sigma_+$,
$\sigma_\times$ and $\omega$, respectively, from (\ref{modeq2})-(\ref{modeq4}), and substituting
in (\ref{modeq1}) followed by one further time differentiation yields, 
\begin{equation}
\ddddot{Z}+2\alpha_2\ddot{Z}+\alpha_1Z=D, \label{mastereq}
\end{equation}
where $\alpha_1=4(p^2+q^2)$, $\alpha_2=\kappa+2k-\beta^2/2$, and $D=-4(pE+qF)$ with $E$ and $F$ as constants of integration. 
Now, one can write the general solution of the ESR variables as,
\begin{equation}
\theta=\frac{\ddot{Z}}{\dot{Z}}-\beta,\;\;\;\;\;\;\;\;
\sigma_+=\frac{pZ+E}{\dot{Z}},\;\;\;\;\;\;\;\; \sigma_\times=\frac{qZ+F}{\dot{Z}},\;\;\;\;\;\;\;\; 
\omega=\frac{G}{\dot{Z}}, \label{esrsolform}
\end{equation}
where $G$ is also a constant of integration, and $Z(t)$ is the solution of the differential equation. This scheme of solution of course
leads to more number of constants of integration than conditions available 
from the initial
values of the ESR variables. Therefore, one has to also ensure the 
satisfaction of the individual
equations in (\ref{thetak})-(\ref{omegak}), which yields appropriate number 
of constraint
equations required for the solution of all the constants of integration 
involved.

It is evident from (\ref{mastereq}) that the solution of $Z(t)$ is dependent 
on the values of $\alpha_1$ and $\alpha_2$,
which are in turn dependent on the curvature, stiffness and damping parameter values. The solutions for
all possible cases are straightforward, though may be tedious.
Therefore, for the purpose of illustration, we consider only some special cases in detail in the
rest of this section. The nature of solutions for all possible cases are discussed and summarized 
separately later. \\

\noindent {\bf Case I}: $k=k_+=k_\times=0$ and $\beta=0$ 

We start with the case without stiffness and damping.
Here, we have some simplifications in (\ref{mastereq}) with 
$\alpha_1=1$ and $\alpha_2=\kappa$. The solutions for the spherical and 
hyperbolic geometries are presented below.\\

\noindent {\em Spherical geometry $(\kappa=1)$}  

The exact solutions for the ESR variables in this case are obtained as
follows. 
The solution for the equation for $Z(t)$ is obtained as:
\begin{equation}
Z(t) = \left ( A_1 + B_2 + B_1 t\right ) \sin t + \left (B_1-A_2-B_2 t\right ) \cos t - 2E .
\end{equation}
From this expression, using the above--mentioned definitions for the ESR
variables we obtain:

\begin{equation}
\theta  = \frac{(A_2 + B_1 + B_2 t) \cos t + (B_2 -A_1 -B_1 t) \sin t}{ (A_1 + B_1 t) \cos t + (A_2 +B_2 t) \sin t},
\label{theta1}
\end{equation}
\begin{equation}
\sigma_{+} = \frac{2 E -\{ (A_1 + B_2 + B_1 t) \sin t + (B_1 -A_2 -B_2 t) \cos t\}}{ 2 [(A_1 + B_1 t) \cos t + (A_2 +B_2 t) \sin t]},
\label{sigma+1}
\end{equation}
\begin{equation}
\sigma_{\times}  = \frac{F}{ [(A_1 + B_1 t) \cos t + (A_2 +B_2 t) \sin t]} ,
\label{sigmacross1}
\end{equation}
\begin{equation}
\omega = \frac{G}{ [(A_1 + B_1 t) \cos t + (A_2 +B_2 t) \sin t]}
.\label{omega1}
\end{equation}
where, $A_1$, $B_1$, $A_2$, $B_2$, $E$, $F$ and $G$ are the integration constants and will also be used throughout in the text for the analytical solutions of other cases as well. These can be defined in terms of the initial conditions on the ESR variables at $t=0$ which are represented as $\theta_0, \sigma_{+0}, \sigma_{\times 0}$ and $\omega_0$. The integration constants are calculated by putting back the solutions (\ref{theta1})-(\ref{omega1}) in equation (\ref{theta})
and using the initial conditions on the ESR variables as, 
$$
A_1 = \frac{4}{(\theta_0^{2}+ 4\sigma_{+0}^2) - 4 I_0},\, \, ~~~~~~~ A_2 = \frac{A_1(\theta_0 + 2\sigma_{+0})}{2},~~~~~ B_1 = \frac{A_1(\theta_0 - 2\sigma_{+0})}{2},\, \,    \nonumber
$$
\begin{equation}  
B_2 = 1,~~~~~~~ E= 0, ~~~~~~~~\, \, \, F = \, A_1 \, \sigma_{\times 0}, \, \, \,~~~~~~~ G=  A_1\,\omega_{0},  \label{ic}
\end{equation} 
where $I_0 = \sigma_{\times0}^2 -\omega_0^{2}$. The constant $B_2$ is arbitrarily chosen as unity.\\

\noindent {\em  Hyperbolic geometry $(\kappa=-1)$}  

In this case, the  exact solutions of the ESR variables can be expressed as, 
\begin{equation}
\theta  = \frac{(A_2 + B_1 + B_2 t) \cosh t + (A_1 + B_2 + B_1 t) \sinh t}{ (A_1 + B_1 t) \cosh t + (A_2 +B_2 t) \sinh t},
\label{theta2}
\end{equation}
\begin{equation}
\sigma_{+} = \frac{2 E -\{ (A_1 - B_2 + B_1 t) \sinh t + (A_2 -B_1+B_2 t) \cosh t\} }{ 2[\,(A_1 + B_1 t) \cosh t + (A_2 +B_2 t) \sinh t]},
\label{sigma+2}
\end{equation}
\begin{equation}
\sigma_{\times}  = \frac{F}{[(A_1 + B_1 t) \cosh t + (A_2 +B_1 t) \sinh t]} ,
\label{sigmacross2}
\end{equation}
\begin{equation}
\omega = \frac{G}{[(A_1 + B_1 t) \cosh t + (A_2 +B_1 t) \sinh t]}
.\label{omega2}
\end{equation}
The integration constants for this case remain same as given by (\ref{ic}).
It may be noted that, in the absence of stiffness and damping, the solutions of the ESR variables
in spherical geometry are
in terms of harmonic functions, while the solutions are composed of hyperbolic 
functions in 
hyperbolic geometry. Therefore, while in spherical geometry a finite time 
singularity is inevitable
(irrespective of initial conditions), in hyperbolic geometry a 
solution singularity may be absent for
specific initial conditions. This fact can be noted by looking at the
values of $t$ for which the denominator can go to zero in each case. For the
spherical geometry the denominator goes to zero if: 
\begin{equation}
\tan t = -\frac{A_1+ B_1 t}{A_2 + B_2 t}.
\end{equation}
The right hand side has value $\frac{A_1}{A_2}$ at $t=0$, and as $t \rightarrow \infty$, we get $-\frac{B_1}{B_2}$. Since the left hand side is not a bounded function there
will always be solutions of the above equation at finite values of $t$. On the
other hand, for a hyperbolic geometry, we have,
\begin{equation}
\tanh t = -\frac{A_1+ B_1 t}{A_2 + B_2 t}.
\end{equation}
The left hand side is now a bounded function (takes on values between $-1$ and $+1$). 
Thus, if $ A_1> A_2 $ and $B_1>B_2$, we find that there are no solutions
which reach a finite time singularity. \\

\noindent {\bf Case II}: $\alpha_2=0$ and $\kappa=\pm 1$  

This case involves both stiffness and damping.
The choice $\alpha_2=0$ implies $2k-\beta^2/2=\mp 1$ (for $\kappa=\pm 1$) which signifies a relation between the damping parameter $\beta$ and the
parameter $k$ related to the elastic force. Hence, one might be tempted
to note a similarity between this case and that of standard critical
damping (say for a harmonic oscillator). 

The solution of equation (\ref{mastereq}) then, consequently 
leads to the following ESR variables,
\begin{equation}
\theta  = \frac{2\, \gamma [  e^{\gamma t}\{ B_1 \cos {\gamma t} - A_1 \sin {\gamma t} \} + e^{-\gamma t} \{A_2 \sin {\gamma t} - B_2 \cos {\gamma t} \}] } {e^{\gamma t} \{D_1 \sin {\gamma t} + C_1 \cos {\gamma t}\} + e^{-\gamma t} \{D_2 \cos {\gamma t}  - C_2 \sin {\gamma t} \} } \ - \beta,
\label{theta21}
\end{equation}
\begin{equation}
\sigma_{+} = \frac{E -(\frac{1}{2} + k_{+})[\frac{S}{\gamma^4} +  e^{\gamma t}\{A_1 \cos{\gamma t} + B_1 \sin {\gamma t}\} +  e^{-\gamma t} \{ A_2 \cos{\gamma t} + B_2 \sin{\gamma t}\}]}{\gamma[ e^{\gamma t} \{D_1 \sin {\gamma t} + C_1 \cos {\gamma t}\} + e^{-\gamma t} \{D_2 \cos {\gamma t}  - C_2 \sin {\gamma t} \}]},
\label{sigma+3}
\end{equation}
\begin{equation}
\sigma_{\times}  = \frac{F - k_\times [\frac{S}{\gamma^4} +  e^{\gamma t} \{A_1 \cos {\gamma t} + B_1 \sin {\gamma t}\} +  e^{-\gamma t} \{ A_2 \cos {\gamma t} + B_2 \sin {\gamma t}\}]}{\gamma[ e^{\gamma t} \{D_1 \sin {\gamma t} + C_1 \cos {\gamma t}\} + e^{-\gamma t} \{D_2 \cos {\gamma t}  - C_2 \sin {\gamma t} \}]} ,
\label{sigmacross3}
\end{equation}
\begin{equation}
\omega = \frac{G}{\gamma[e^{\gamma t} \{D_1 \sin {\gamma t} + C_1 \cos {\gamma t}\} + 
e^{-\gamma t} \{D_2 \cos {\gamma t}  - C_2 \sin {\gamma t} \}]} ,\label{omega11}
\end{equation} 
where $\gamma=\alpha_1^{1/4}$ and $C_1 = A_1+B_1$, $D_1 =B_1-A_1 $, $C_2 = A_2+B_2$, $D_2 = B_2-A_2$, 
$E_1$, $E_2$ and $G_1$ are the integration constants. These constants are calculated by putting 
back the solutions in the corresponding evolution equations and using the initial conditions of
the ESR variables as follows,
\begin{eqnarray}
&&A_1 = \frac{1}{4}(2 P -Q +1),\qquad B_1=\frac{1}{4}(2P +Q-1) \nonumber , \\
&&A_2=\frac{1}{4\gamma}(\theta_0  + \gamma Q + \gamma), \qquad B_2 = \frac{1}{4 \gamma}({-\theta_0} + {\gamma} Q + {\gamma}), \nonumber\\
&& 
E = \gamma \, \sigma_{+0},\qquad F=  \gamma \, \sigma_{\times 0},  \qquad G = \gamma \, \omega_0,
\end{eqnarray}
where $P$ and $Q$ are given as follows,
\begin{eqnarray}
 &&  P =  - \frac{1}{2\gamma^3}\, [(1 + 2 k_+)\sigma_{+0} -2 k_\times \sigma_{\times 0}], \qquad Q = \frac{1}{2 \gamma^2}\, [{\theta_0^2} - 2(\sigma_{+0}^2 + \sigma_{\times 0}^2 -\omega_0^{2})].
\end{eqnarray}
In calculating the integration constants $E$ and $F$, we have further used $Z(0) = 0$ without any loss of 
generality. It may be noted that the solution of $Z(t)$ consists of one exponentially growing and the other 
exponentially decaying harmonic mode. Thus, there is a finite time at which $\dot{Z}(t)=0$ irrespective
of the initial conditions. This happens when the denominator in the
above equations goes to zero implying,
\begin{equation}
\tan \gamma t = \frac{C_1 e^{\gamma t} + D_2 e^{-\gamma t}}{C_2 e^{-\gamma t}
-D_1 e^{\gamma t}}.
\end{equation}
The fact that the above condition can be satisfied for some values of $t$
confirms that the ESR variables can have finite 
time singularities. It may also be noted that a difference in
the above-mentioned
forms of the ESR variables for the $\kappa=1$ and
$\kappa=-1$ cases arise only through the relation between the allowed
values of $k$ and $\beta$ (i.e. $2k-\beta^2/2=\mp 1$ for $\kappa=\pm 1$). Only the value of the additive constant piece $\beta$ changes in the
functional form of $\theta$, when we change $\kappa$ from
$+1$ to $-1$.   
The various generic features based on the analytical solutions for all possible cases are 
presented in the next section.

\subsection{Nature of solutions}
\begin{figure}
\centerline
\centerline{\includegraphics[scale=0.7]{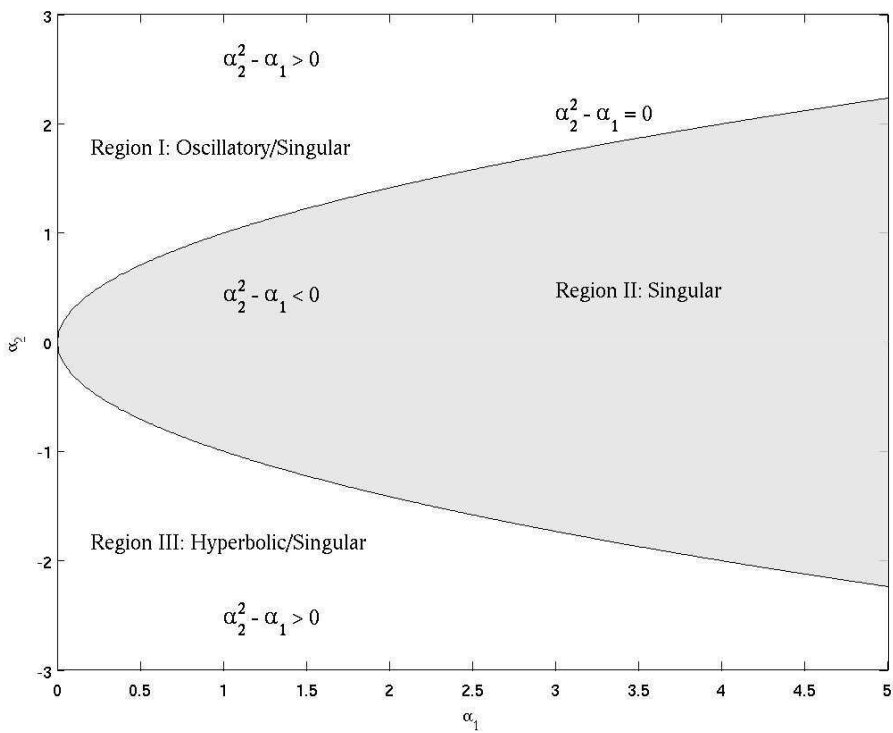}}
\vspace{-4.0cm}
\caption{Identification of different regions in the $\alpha_1$-$\alpha_2$ parameter plane
corresponding to different generic features of the solution of (\ref{mastereq}).}
\label{fig1}
\end{figure}
The generic features of the solutions of (\ref{mastereq}) for different values of 
$\alpha_1$ and $\alpha_2$ are straightforward to obtain from 
the original
fourth order differential equation. To see this, substitute
$Z(t)=Y(t)+ D/\alpha_1$ to get rid of the constant inhomogeneous term $D$. The equation for
$Y(t)$ can now be easily solved by the substitution $Y(t) = e^{i r t}$. We obtain the relation $r^4 - 2\alpha_2 r^2 +\alpha_1 =0$ from which, by solving, we get:
\begin{equation}
r = \pm \left (\alpha_2 \pm \sqrt{\alpha_2^2-\alpha_1} \right )^{1/2}.
\end{equation}
Thus, the nature of the solutions crucially depends on the value of
$\alpha_2^2-\alpha_1$, the sign of $\alpha_2$ and also on whether
$\alpha_1>0$ or $\alpha_1=0$.

The nature of possible solutions 
is schematically presented in the $\alpha_1$-$\alpha_2$ parameter
plane in Fig.~\ref{fig1}. Based on this figure, one can comment on the
nature of solutions of the ESR variables, as summarised in Table \ref{gamma}. 

\begin{table}
\small
\begin{center}
\begin{tabular}{|l|l|}
\hline\, \, \,  Conditions \, \, \, & \, \, \,\, \, \,\, ~~~~~~~~~~ Nature of Solutions \\
 \hline 
\,$\alpha_2^2 >\alpha_1 $&\\
 \, $(i) \, \alpha_2>0$: Region I&\\
\hspace{0.1in} \,\, \, $\bullet\, \,  \alpha_1>0$ \, \, \, & Finite time singularity irrespective of initial conditions  \\
\hspace{0.1in} \,\, \, $\bullet \, \, \alpha_1=0$ \, \, \, & Initial condition dependent oscillatory solution, or finite time singularity \\
\,$ (ii) \,  \alpha_2<0$: Region III&\\
\hspace{0.1in} \,\, \, $\bullet \, \, \alpha_1>0$ \, \, \, & Hyperbolic solutions: initial condition dependent stable solution, \\
& or finite time singularity \\
\hspace{0.1in} \,\, \, $\bullet  \, \, \alpha_1=0$ \, \, \, &  Initial condition dependent stable solution, or finite time singularity  \\
 \hline \,  $\alpha_2^2=\alpha_1$ & Finite time singularity irrespective of initial conditions\\ \hline  \, $\alpha_2^2<\alpha_1$: Region II\, &\,Oscillatory solutions with one growing and one decaying mode \\ \hline
\end{tabular}
\end{center}
\caption {Different possible nature of solutions of the ESR variables corresponding to
 different regions as marked in Fig. \ref{fig1}.} 
\label{gamma}
\end{table}
It is worth noticing from the expression of $\alpha_2=\kappa+2k-\beta^2/2$ that a large value of $k$ in hyperbolic geometry can make the solutions similar to 
that in spherical geometry. This will
occur for values of $k$ such that $\alpha_2^2>\alpha_1>0$. On the other hand,
a large value of $\beta$ such that $\alpha_2^2>\alpha_1$ and $\alpha_2<0$, 
in a spherical geometry the flow characteristics will resemble those in 
a hyperbolic geometry. Further, with suitable values of stiffness and
damping, the evolution of the ESR variables in a curved space (spherical or hyperbolic) can exhibit
features of evolution in a flat (Euclidean) space without stiffness and damping. This will occur
whenever $\alpha_2=0$, $k_+=-1/2$ and $k_\times=0$.

The nature of solutions depending on initial conditions can also be easily determined. As a specific
interesting example, consider the case with $\alpha_1=0$ and $\alpha_2>0$. In this case (\ref{mastereq})
reduces to, 
\begin{equation}
\ddddot{Z}+2\alpha_2\ddot{Z}=0, \nonumber 
\end{equation}
which has the general solution,
\begin{equation}
Z(t)=A_1\cos\sqrt{2\alpha_2}t +B_1\cos\sqrt{2\alpha_2}t +A_2t +B_2,
\end{equation}
where $A_1$, $B_1$, $A_2$ and $B_2$ are constants of integration, which are determined
from the initial conditions. Now, as is evident from
(\ref{esrsolform}), the ESR variables will exhibit finite time singularity whenever $\dot{Z}=0$.
This will occur whenever $|A_2|<\sqrt{2\alpha_2(A_1^2+B_1^2)}$, which can now be explicitly
written in terms of the initial conditions on the ESR variables. It may be noted that
if there is no singularity, the ESR solution in this case will exhibit stable oscillations.

\subsection{Numerical examples}\label{visu}
In this section, we numerically investigate the nature of solutions 
for different initial conditions and parameter values 
($\alpha_1$ and $\alpha_2$) in different regions as
indicated in Fig. \ref{fig1}. 

\begin{enumerate}
\begin{figure}[H]
\centerline
\centerline{\includegraphics[scale=0.7]{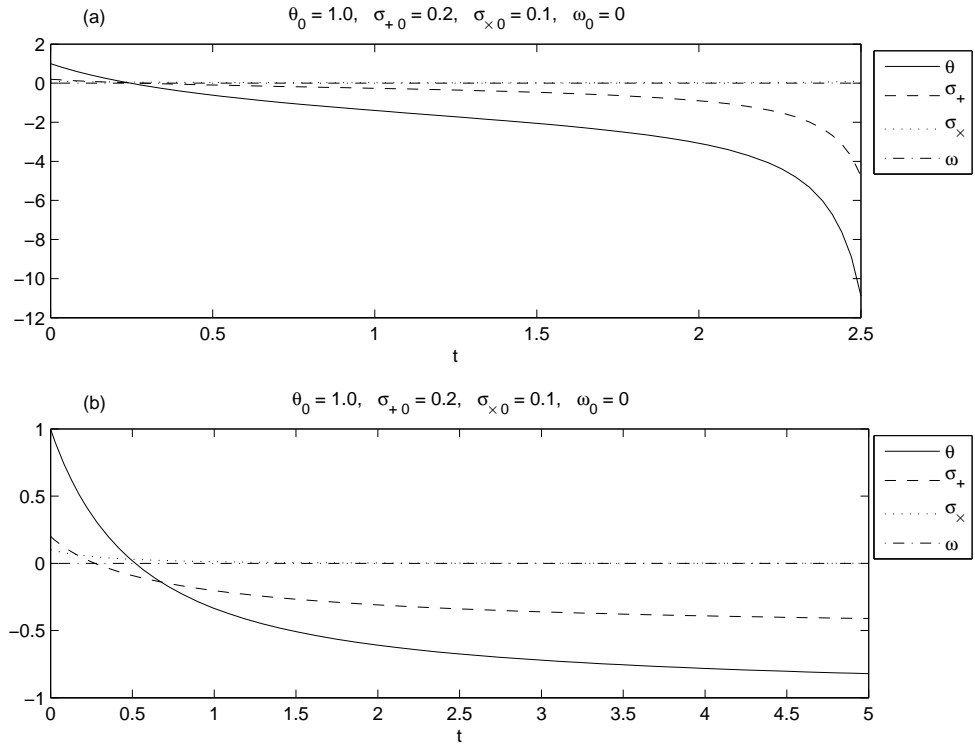}}
\caption{Evolution of ESR variables with parameters $k=1.0$, $k_+ = k_\times = 0$, $\beta=2.0$,
for (a) $\kappa=1$, and (b) $\kappa=-1$.}
\label{fig2}
\end{figure}
\item{$\alpha_1>0$ and $\alpha_2^2=\alpha_1$\\
These parameter values lie exactly on the parabola in Fig.~\ref{fig1}. When $\alpha_2>0$, the solutions
of $Z(t)$ are composed of harmonic functions with secular terms. This makes the solution of the ESR
variables singular in finite time irrespective of the initial conditions, as shown in Fig.~\ref{fig2}(a). 
On the other hand, when $\alpha_2<0$, the
solutions of $Z(t)$ have hyperbolic functions with secular terms. In this case, it is possible to 
find initial conditions on the ESR variables for which the solutions are non-singular. 
One such case is shown in Fig.~\ref{fig2}(b). 
}
\item{$\alpha_1=0$ and $\alpha_2^2>0$\\
In this case, the parameter values lie on the positive $y$-axis in Fig.~\ref{fig1}.
As discussed before, the ESR solutions can show finite time singularity or pure oscillatory
modes, depending on the initial conditions. Both these cases are presented in Fig.~\ref{fig3}(a)
and (b). It is interesting to note that the difference in initial conditions in the two
sub-plots is only in the rotation initial condition (i.e., $\omega_0$). The
stabilizing effects of rotation (also observed in \cite{adg}) is evident once again.} 
\begin{figure}[H]
\centerline
\centerline{\includegraphics[scale=0.7]{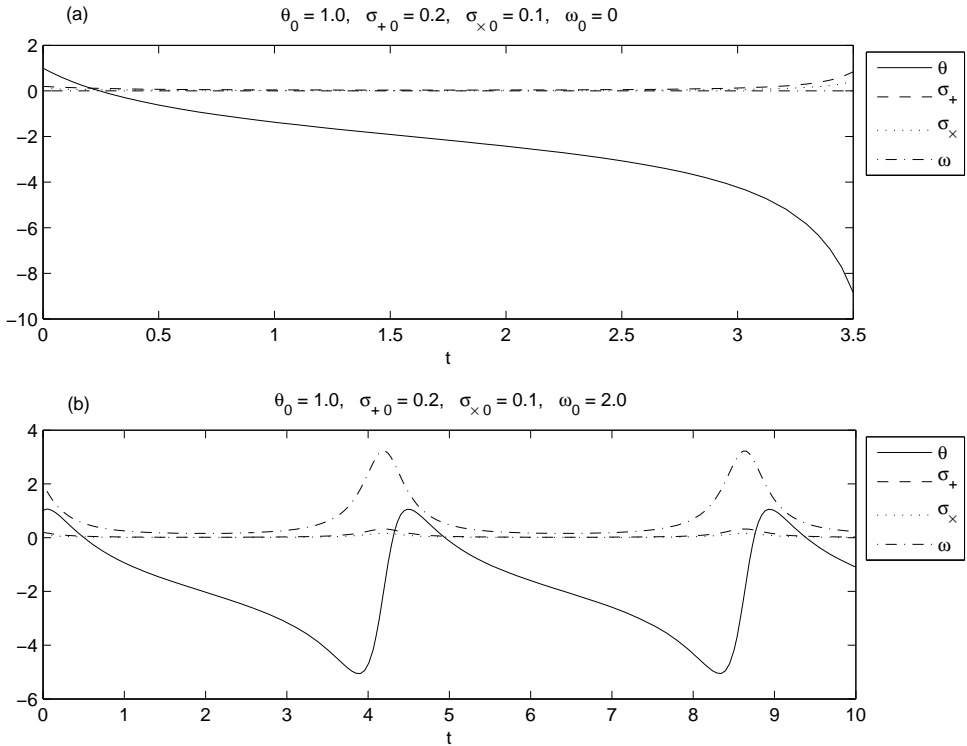}}
\caption{Evolution of ESR variables with parameters $\kappa=1$, $k=1.0$, $k_+=-0.5$, $k_\times=0$, $\beta=2.0$
for two different initial conditions.}
\label{fig3}
\end{figure}

\begin{figure}[H]
\centerline
\centerline{\includegraphics[scale=0.7]{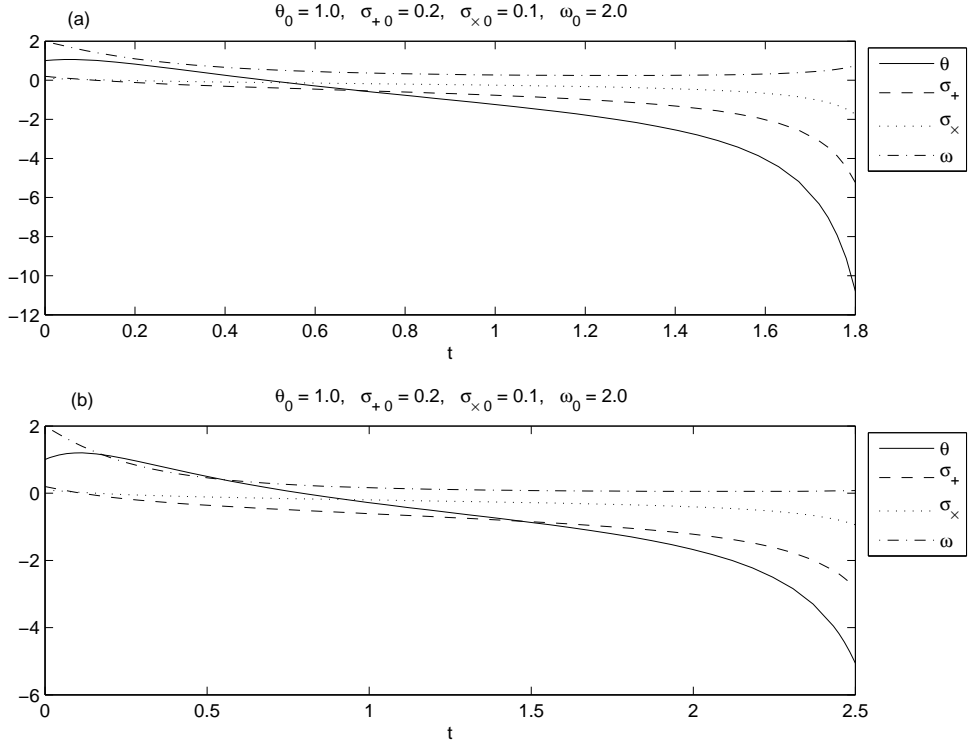}}
\caption{Evolution of ESR variables with parameters $k=1.0$, $k_+=1.0$, $k_\times=0.5$, $\beta=2.0$
for (a) $\kappa=1$ and (b) $\kappa=-1$, both of which lie in Region II in Fig.~\ref{fig1}.}
\label{fig4}
\end{figure}

\item{$\alpha_1>0$ and $\alpha_2^2<\alpha_1$\\

These parameter values lie in Region II in Fig.~\ref{fig1}. Hence, they will all show finite time
singularity as discussed before.

 Two such cases have been presented in Figs.~\ref{fig4}(a) and (b) for
spherical and hyperbolic geometries, respectively.}

\item{$\alpha_1>0$ and $\alpha_2^2>\alpha_1$
\begin{figure}[H]
\centerline
\centerline{\includegraphics[scale=0.69]{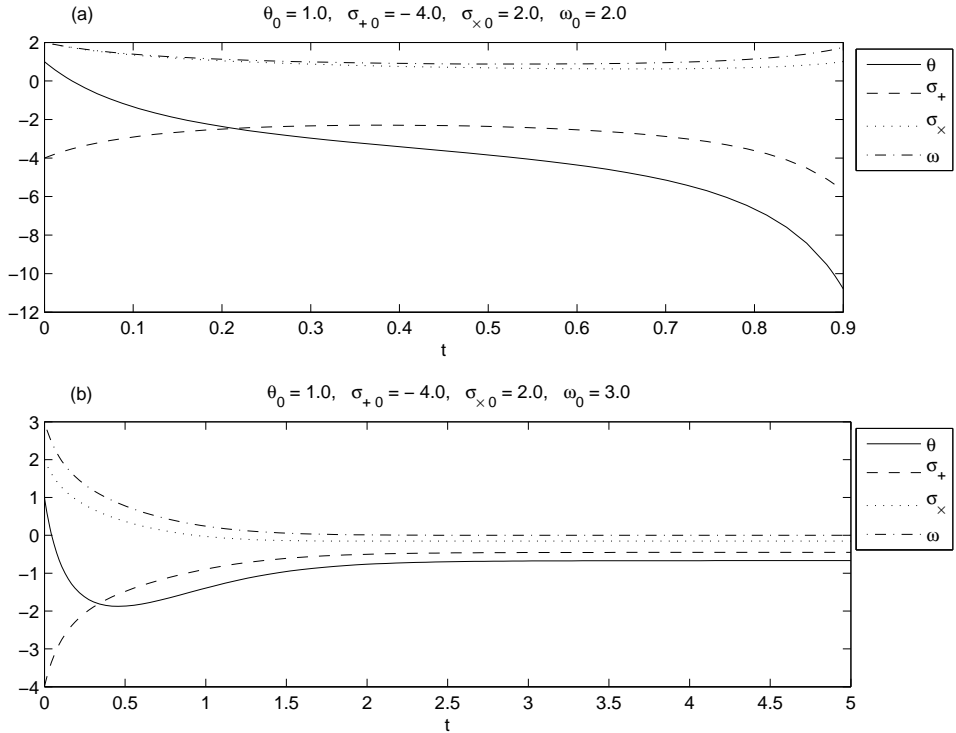}}
\caption{Dependence of the nature of solutions of the ESR variables for two different sets of initial
conditions with $\kappa=-1$, $k=0.5$, $k_+=1.0$, $k_\times=0.5$, $\beta=4.0$.}
\label{fig5}
\end{figure}}
\end{enumerate}
We consider here parameter values in Region III in Fig.~\ref{fig1}. The solutions of $Z(t)$ here
are composed of hyperbolic functions. The nature of solutions of the ESR variables are now
dependent on the initial conditions. One can, therefore, have finite time singularity, as
shown in Fig.~\ref{fig5}(a), or stable solutions, as observed in Fig.~\ref{fig5}(b).

\section{Geodesic flows on a torus}
In this section, we consider kinematics of deformations on a torus which represents a curved medium with varying curvature. 
The line element for a  ring torus is given as,
\begin{equation}
ds^2 = (1 + a \cos \phi)^2 d\psi^2  \,  +  a^2 d \phi^2 ,
\label{metrictori}\end{equation}
where $\psi, \phi \rightarrow [0, 2 \pi]$ and  $ a = r_1 / r_2 $ with $r_1$ and   $r_2$ as the minor and  major radius of the torus, respectively.  
\subsubsection{Raychaudhuri  equations}
Using (\ref{dotb}) and (\ref{metrictori}), the Raychaudhuri equations for ESR variables can now be obtained as follows,
\begin{equation}
\dot \theta + \frac{1}{2}{\theta^2}+\beta \theta  + 2 (\sigma_{+}^{2} + \sigma_{\times}^{2} -\omega^2) + \frac{\cos \phi}{a(1+a\cos \phi)}  + 2k =0,
\label{thetat}
\end{equation}
\begin{equation}
\dot \sigma_{+}+ (\beta +\theta) \, \sigma_{+} - \frac{\cos \phi}{2a (1+a\cos \phi)}  + k_{+} =0,
\label{sigma+t}
\end{equation}
\begin{equation}
\dot \sigma_{\times}+ (\beta +\theta) \, \sigma_{\times}  + k_{\times}=0, 
\label{sigmacross0t}
\end{equation}
\begin{equation}
\dot \omega+ (\beta + \theta)\, \omega  =0
.\label{omegat}
\end{equation}
One can try to solve these equations by simultaneously solving the geodesic equations corresponding to the metric (\ref{metrictori}).

\subsubsection{Geodesic equations}
The geodesic equations corresponding to the above metric are given by,
\begin{equation}
\ddot \psi - \frac{2 a \sin \phi}{ (1 + a \cos \phi) } \, \dot \psi\, \dot \phi = 0,\label{psi1}     
\end{equation}
\begin{equation}
\ddot \phi + \frac{(1 + a \cos \phi) \sin \phi }{a}\, \dot \psi^2  = 0. 
\label{phi1}
\end{equation}
The first integrals of these geodesic equations are obtained as,
\begin{equation}
\dot \psi = \frac{l_1}{(1 + a \cos \phi)^2}, 
\label{tpsi1}    
\end{equation}
\begin{equation}
\dot \phi = \sqrt{\frac{-l_1^2}{a^2(1 + a \cos \phi)^2} + l_2}, 
\label{tphi1}
\end{equation}
where $l_1$ and $l_2$ are integration constants and their different choices lead to five different families of geodesics \cite{Mark}.
 The normalisation $g_{\alpha \beta} u^\alpha u^\beta = 1$ for this case yields $l_2= 1/a^2$, and the equation (\ref{phi1}) is therefore only dependent 
on $l_1$. We now consider some special cases in order to perform a detailed 
analysis of the 
deformations as obtained by integrating the Raychaudhuri equations (\ref{thetat})-(\ref{omegat}).
  
\subsubsection{Kinematics of ESR for certain geodesic families}

We comment here on the nature of solutions of Raychaudhuri equations  for 
certain geodesics families. 

\noindent {\bf Case I}: $\dot \phi =0$ and $\dot \psi \neq 0$ \\
This case corresponds to the inner equator geodesic of a torus. From equations  (\ref{tpsi1}) and (\ref{tphi1}), we have  $\phi =$ constant and $\dot \psi = 1/l_1$.
\vspace{-0.5cm}
\begin{figure}[H]
\centerline{\includegraphics[scale=0.8]{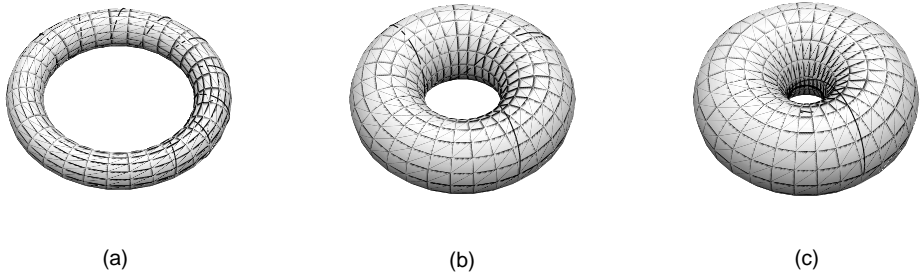}}
\caption{Geodesics  on a torus considered for the numerical integration of the Raychaudhuri equations for (a) $a=0.2$, (b) $a=0.5$, and (c) $a=0.7.$}
\label{torusfig}
\end{figure}
 The Raychaudhuri equations are given by (\ref{thetak})-(\ref{omegak}) with the substitutions  $\kappa = \gamma =  {\cos \phi} / {a(1+a\cos \phi)}$ (which is constant in this case), $p = \gamma/2 -k_+$ and $q =-k_\times$.  The value of $\gamma$ can be positive, negative or zero. For positive (negative) value of $\gamma$, the solution can be obtained from the spherical (hyperbolic) solution as presented in the previous section. On the other hand, for $\gamma =0$ (corresponding to $\phi = \pi / 2 , \, 3 \pi / 2$), the solution follows from the flat space solutions as obtained in \cite{adg}. Therefore, the kinematics of deformations are similar to the cases discussed previously. \\           
\noindent {\bf Case II}: $\dot \phi \neq 0$ and $\dot \psi =  0$ \\
This case corresponds to the meridians of a torus. From equations (\ref{tpsi1})
and (\ref{tphi1}), we now have  $\dot \phi = 1/a$ and $\psi =$ constant. The Raychaudhuri equations have the same structure as in Case I above with $\gamma = \gamma (\lambda)$. It is difficult to integrate these equations 
analytically because of the variation of $\gamma $ with $\lambda$. 
However, we can integrate them numerically.  

\begin{figure}[H]
\centerline{\includegraphics[scale=0.7]{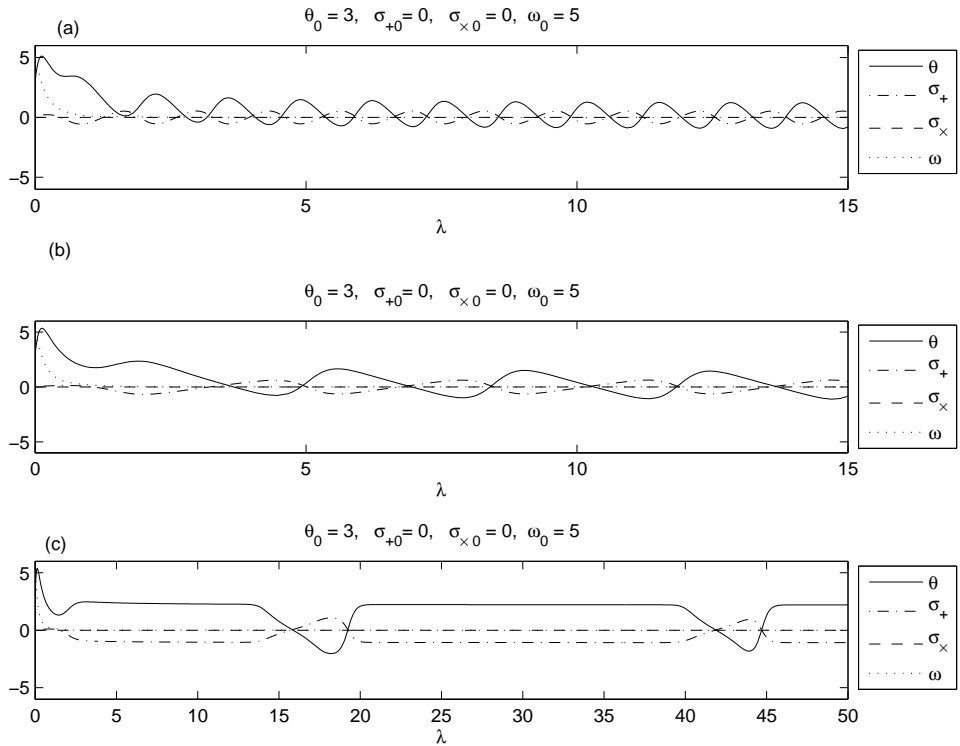}}
\caption{Evolution of ESR variables for different values of $a$ corresponding to  the three geodesics shown in Fig. \ref{torusfig} 
($l_1=0.3$, $l_2=1/a^2$).}
\label{torusfig2}
\end{figure}

We now present numerical solutions of the Raychaudhuri equations corresponding 
to different values of the dimensionless parameter $a$. The corresponding 
geodesics are shown in Fig \ref{torusfig} which belong to two geodesic 
families on a torus \cite{Mark}. Excluding the geodesics corresponding to 
the above mentioned Case I, the evolution of the ESR variables is observed 
to be oscillatory. These oscillations are basically due to periodic variations 
of $\gamma = R/2$ (where $R$ is Ricci scalar) with $\phi$.  The frequency and 
amplitude of oscillations for small values of $a$ is high and reduces with the 
increase in $a$. The evolution of the ESR variables is presented in Fig. \ref{torusfig2} for three different values of $a$. Solutions may
or may not exhibit a finite time singularity, depending on the initial values of $\theta_0$ and $\omega_0$.

The Raychaudhuri equations for other known families of geodesics on the torus 
are, similarly, difficult to solve analytically and can be studied numerically.
We do not indulge in discussing them further in order to avoid
repeating qualitatively similar conclusions. 

\section{Comment on evolution  equations in three dimensional media}
The general expression for the evolution tensor $B_{ij}$ in in three dimensional media \cite{adg} is expressed as:
\begin{equation}
B_{ij} = \frac{1}{3}{\theta} \, \,\delta^i_{\,\,j}  + \sigma^i_{\,\,j} + \omega^i_{\,\,j} , \label{bij3d}
\end{equation}
where the expansion $(\theta)$, shear $(\sigma^i_{\,\,j} )$  and rotation $(\omega^i_{\,\,j})$ in the equation (\ref{bij3d}) have the following $3\times 3$ matrix form :
\begin{equation}
B^i_{j} =
\left(
\begin{array}{ccc}
\frac{1}{3}\theta & 0&0\\
0&  \frac{1}{3}\theta&0\\
0&0&\frac{1}{3}\theta\\
\end{array} \right)
 + \left(
\begin{array}{ccc}
\sigma_{11} & \sigma_{12}&  \sigma_{13} \\
\sigma_{12} & \sigma_{22}&  \sigma_{23}\\
 \sigma_{13} & \sigma_{23}&-(\sigma_{11} +\sigma_{22})\\
\end{array} \right)
 +
\left(
\begin{array}{ccc}
0 & -\omega_3 & \omega_2 \\
\omega_3& 0 & -\omega_1\\
-\omega_2&\omega_1&0\\
\end{array} \right).
 \label{ctrans1}
\end{equation}
 It may be noted that, for simplicity, we have used the definition $\sigma^i_{\,\,j} = \sigma_{ij}$ and $\omega^1_{\,\,2}=-\omega_3$, $\omega^1_{\,\,3}= \omega_2$ and $\omega^2_{\,\,3}=-\omega_1$. In order to derive the evolution equations in three dimensions 
let us consider, the following three dimensional line element,
\begin{equation}
ds^2 = \frac{ dr^2}{ 1-\kappa r^2} \,  +  r^2 (d \varphi^2 + \sin ^2\varphi\, d\chi^2) .
\label{metric1}
\end{equation}
This line element represents three dimensional maximally symmetric
spaces of constant curvature which are topologically $S^3$ ($\kappa=1$),
$H^3$ ($\kappa=-1$) or $R^3$ ($\kappa=0$).

\noindent The evolution equations (i.e. Raychaudhuri equations) for the
expansion, shear and rotation can be derived following standard methods.
There are nine coupled, first order, nonlinear equations 
involving $\theta$ (expansion), $\sigma_{11}, \, \sigma_{12},\,\sigma_{22}\,,\sigma_{13}$,\,$\sigma_{23}$ (shear) and $\omega_{1},\,\omega_{2},\,\omega_{3}$ (rotation) and their first derivatives. 

The geodesic equations in the above mentioned line element  
are as follows,
\begin{equation}
\ddot r + \frac{\kappa r}{(1-\kappa r^2)} \dot r^2 - r\, (1-\kappa r^2)\,  \dot \varphi ^2 - r\, (1-\kappa r^2)\,  \sin^2 \varphi \, \dot \chi ^2 = 0,
\label{rg12}
\end{equation}
\begin{equation}
r \, \ddot \varphi + {2 \dot r}\, \dot \varphi - r \sin \varphi\,  \cos \varphi \, \dot \chi ^2  =0,
\label{tg12}
\end{equation} 
\begin{equation}
r \, \ddot \chi + {2 \dot r} \, \dot \chi + 2 \cot \varphi \, \dot \varphi\,  \dot \chi =0.
\label{cg}
\end{equation} 
For the three dimensional case, it may be noted 
that the equations (\ref{rg}) and (\ref{tg}) are modified with additional terms and we also have a third equation (\ref{cg}). These equations reduce 
to the geodesic equations for the two dimensional case 
for a constant value of $\chi$.

We make a particular choice of geodesics for which
$\chi$= constant, $\dot{\phi}=d$, $r=1$ and with $\kappa =1$ (i.e. the case of 
spherical geometry). Further, $K_{ij}=k \delta_{ij}$. It is also
possible to consistently choose $\sigma_{13} =
\sigma_{23} =0, \omega_1=\omega_2=0$. With all of these choices,
the evolution equations for the five remaining variables turn out to be,
\begin{equation}
\dot \theta + \frac{1}{3}{\theta^2}+\beta \theta  + 2 (\sigma_{11}^{2} +\sigma_{22}^{2} + \sigma_{12}^{2} +\sigma_{11}\sigma_{22} -\omega_3^{2}) +3k + 6 =0
,\label{theta31}
\end{equation}
\begin{equation}
\dot \sigma_{11}+ (\beta + \frac{2}{3}\theta) \, \sigma_{11} + \frac{1}{3} (\sigma_{11}^{2}  + \sigma_{12}^{2} - \omega_3^{2})  - \frac{2}{3} ( \sigma_{22}^{2}+\sigma_{11}\sigma_{22}) -2  = 0
,\label{sigma111}
\end{equation}
\begin{equation}
\dot \sigma_{12}+ (\beta + \frac{2}{3}\theta + \sigma_{11} + \sigma_{22}) \, \sigma_{12}  =0
,\label{sigma121}
\end{equation}
\begin{equation}
\dot \sigma_{22}+ (\beta + \frac{2}{3}\theta) \, \sigma_{22} + \frac{1}{3} (\sigma_{21}^{2}  + \sigma_{22}^{2} - \omega_3^{2})  - \frac{2}{3} ( \sigma_{11}^{2}+\sigma_{11}\sigma_{22}) - 2 = 0
,\label{sigma221}
\end{equation}
\begin{equation}
\dot \omega_3 + (\beta + \frac{2}{3} \theta + \sigma_{11} + \sigma_{22}) \, \omega_{3}=0
.\label{omega31}
\end{equation}     
These equations, as well as more general cases can indeed be solved 
numerically.
We have worked through some examples (which we do not quote here). Generic
features are not drastically different. 

\section{Summary and scope}
In this article, we have investigated the kinematics of flows on
curved, deformable media by
solving the evolution equations for the
expansion, shear and rotation of
geodesic congruences in two dimensional spaces of constant as well as
varying
curvature. Let us briefly point out the novelties that arise while
considering flows on curved spaces.
\begin{itemize}
\item{In \cite{adg}, where we looked at flows on flat space, geodesics
were straight lines and the need for solving the geodesic equations did
not arise. Here, however, this is a necessity and we have looked at 
some simple solutions of the geodesic equations which are required as inputs
in the evolution equations for ESR.} 

\item{In a flat space setting, the evolution equations do not
have a contribution from geometric terms involving Ricci and Weyl.
Here, though any contribution from the Weyl term is absent (two
dimensions !), the Ricci term does indeed contribute. Thus, curvature
effects on the nature of the flows do 
appear through the non-zero value of $\kappa$.}

\item{ The analytic solutions obtained for the special cases
illustrate the possibilities of the existence of finite time singularities
in the geodesic congruence under study. We are able to point out,
in a fairly general setting,
the conditions on the parameters which imply the existence of
finite time singularities/nonsingular solutions. The representation
of regions in parameter space where singular/non-singular solutions
exist is also obtained and explained.}

\item{A major difference between flows in positive and negative
curvature spaces is the fact, that in the latter, we do have the
possibility of exclusively non--singular solutions. This is
illustrated through the analytic solutions as well as the
numerical results where we explicitly show the differences
that arise between results in spaces of positive and negative
curvature.}

\item{ The stabilizing effect of rotation, as observed in the
flat space analysis in \cite{adg} persists. We have illustrated
this in our numerical solutions and plots thereof.}
   
\item{The role of stiffness, damping and the initial conditions
are pointed out in each case with representative plots.}

\item{In the case of flows on the torus (varying curvature)
the oscillatory nature of the inhomogeneous term in the Raychaudhuri
equations leads to oscillations in the ESR variables. The nature
of the oscillations (i.e. amplitude and frequency) seem to
depend on the parameter $a$. Further, the solutions exhibit 
finite time singularities and the role of rotation is expected to
be similar, i.e., it can inhibit the formation of finite time
singularities.}

\end{itemize}

Finally, it is customary to ask in conclusion--what next? We have mentioned
briefly the equations in three dimensional spaces of constant 
curvature and pointed out that the results will be similar to those 
in two dimensions. It will surely be of greater value if a complete
D-dimensional analysis is carried out.
In two dimensions, it will be useful to look at further examples of
flows in spaces with varying curvature-- in particular we can look
at higher genus tori and flows on them. 
Furthermore, a case of physical interest could be the kinematics of 
geodesic flows in black holes spacetimes, where the spaces are not
usually of  
constant curvature and also involve a Lorentzian signature line element.

We hope to report on some of the above questions in the near future.

\section*{Acknowledgments}
The authors thank the Department of Science and Technology (DST), Government of India for financial support through 
a sponsored project (grant number: SR/S2/HEP-10/2005).

\end{document}